\documentclass[pra,aps,twocolumn,showpacs,superscriptaddress,unsortedaddress]{revtex4}
\usepackage{graphicx}
\usepackage{amsmath,amssymb}
\usepackage{bm}
\usepackage{psfrag}

\newcommand{\rmi}{\mathrm{i}}
\newcommand{\rmd}{\mathrm{d}}
\renewcommand{\Im}{\mathrm{Im}}

\newcommand{\lpar}{l\,{\parallel}\,l}  
\newcommand{\lperp}{l\,{\perp}\,l}      
\newcommand{\hpar}{h\,{\parallel}\,h}	
\newcommand{\hperp}{h\,{\perp}\,h}      

\newcommand{\ket}[1]{|#1\rangle}
\newcommand{\bra}[1]{\langle#1|}

\newcommand{\eq}[1]{\begin{equation} #1 \end{equation}}
\newcommand{\eqlab}[2]{\begin{equation} #1
	\label{#2.eq}\end{equation}}
\newcommand{\refig}[1]{Fig.~\textup{\ref{#1.fig}}}
\newcommand{\refeq}[1]{\textup{(\ref{#1.eq})}}
\newcommand{\Fe}{F_e}

\newcommand{\Fg}{F_g}
\newcommand{\Ff}{F_f}

\newcommand{\mg}{m_g}

\newcommand{\Pg}{\hat P_g}
\newcommand{\Pe}{\hat P_e}
\newcommand{\bF}{\bm{F}}
\newcommand{\bI}{\bm{I}}
\newcommand{\bJ}{\bm{J}}
\newcommand{\bL}{\bm{L}}
\newcommand{\bS}{\bm{S}}
\newcommand{\sixj}[6]{\left\{\begin{array}{ccc}
		#1	& #2	& #3\\
		#4	& #5	& #6
		\end{array}\right\}}

\newcommand{\journal}[4]{#1 \textbf{#2}, #3 (#4)}

\newcommand{\PRL}[3]{\journal{Phys.\ Rev.\ Lett.}{#1}{#2}{#3}}
\newcommand{\PRA}[3]{\journal{Phys.\ Rev.\ A}{#1}{#2}{#3}}
\newcommand{\PRB}[3]{\journal{Phys.\ Rev.\ B}{#1}{#2}{#3}}
\newcommand{\PRE}[3]{\journal{Phys.\ Rev.\ E}{#1}{#2}{#3}}
\newcommand{\JPA}[3]{\journal{J.\ Phys.\ A: Math.\ Gen.}{#1}{#2}{#3}}
\begin{document}
\author{Cord A.\ M\"uller}
\affiliation{Physikalisches Institut, Universit\"at Bayreuth, D-95440
Bayreuth, Germany}
\author{Christian Miniatura}
\affiliation{Institut Non Lin\'eaire de Nice Sophia Antipolis, UMR 6618 du
CNRS, 1361 route des Lucioles, F-06560 Valbonne, France}
\author{David Wilkowski}
\affiliation{Institut Non Lin\'eaire de Nice Sophia Antipolis, UMR 6618 du
CNRS, 1361 route des Lucioles, F-06560 Valbonne, France}
\author{Robin Kaiser}
\affiliation{Institut Non Lin\'eaire de Nice Sophia Antipolis, UMR 6618 du
CNRS, 1361 route des Lucioles, F-06560 Valbonne, France}
\author{Dominique Delande}
\affiliation{Laboratoire Kastler Brossel, Universit\'e Pierre et Marie Curie, 4 Place Jussieu, F-75005 Paris, France}

\title{Multiple scattering of photons by atomic hyperfine
multiplets}

\date{\today}

\begin{abstract}

Mesoscopic interference effects in multiple scattering of photons 
depend crucially on the internal structure of the scatterers. 
In the present article, we develop the analytical theory of multiple
photon scattering by cold atoms with arbitrary internal hyperfine multiplets. 
For a specific application, we calculate the enhancement factor
of elastic coherent backscattering  as a function of detuning from an entire hyperfine multiplet of
neighboring resonances that cannot be considered isolated.  
Our theory permits to understand why atoms 
behave differently from classical Rayleigh point-dipole
scatterers, and how the classical description is recovered 
for larger but still microscopic objects like molecules or clusters.

\end{abstract}


\maketitle

\section{Introduction}

Atomic physics and quantum optics currently discover the fascinating
field of mesoscopic physics \cite{Houches95}. Mesoscopic phenomena are
due to interference effects that survive a disorder average in
phase-coherent samples. Interference effects can be observed using both genuine quantum 
matter waves such as electrons or ultra-cold atoms and classical waves
such as acoustic or electromagnetic waves 
\cite{AkkerMon_book}.  
Arguably, the most dramatic effect 
is the breakdown of diffusive transport due to  
strong localization, a phenomenon invoked by Anderson in the
context of the metal-insulator transition 
\cite{localizationreviews}.  
Strong localisation has been observed unambiguously
for electrons \cite{Khavin98} and microwaves 
\cite{Chabanov01a} in quasi-one-dimensional systems. 
Yet even in dilute samples far from the dense regime where strong
localisation could be expected, interference effects can be measured.  
In optics, a rather robust interference phenomenon is the coherent backscattering
(CBS) effect. Here, the constructive interference between waves
counterpropagating along a given multiple scattering path enhances
the average diffuse intensity reflected from an ensemble of random 
samples in a narrow angular range around the
backscattering direction 
\cite{CBS:Exp}. 
Under optimal experimental conditions allowing to apply 
the reciprocity theorem \cite{reciprocity}, 
this two-wave interference enhances the intensity 
exactly by a factor of 2 \cite{wiersma95}. 

Quantum optical systems involving multiple scattering of photons
by atoms are well adapted to study general concepts of quantum
transport. 
Indeed, atoms are very efficient point scatterers for light  
because the scattering cross section close to an
internal resonance is huge compared to the actual atomic size. 
Laser-cooling techniques permit one to prepare low temperature clouds 
where a negligible Doppler broadening 
of atomic transitions preserves the phase coherence of
the propagating photons. 
For low enough laser intensity, photons are scattered completely
elastically from closed atomic dipole transitions such that inelastic
scattering and absorption are absent (see, however, \cite{saturation} for 
saturation effects at higher field intensities). 
Furthermore, by injecting and detecting well-defined
photon polarization states, one is capable of probing the internal
spin degrees of freedom of the atomic scatterers.  

The experimental observation of CBS of laser
light by cold atoms \cite{CBS:At,Kupriyanov} has
revealed the crucial impact of the internal atomic structure onto coherent photon transport. Atomic
dipole transitions with a nondegenerate ground-state scatter photons 
like isotropic point-dipoles (also known as Rayleigh
scatterers) \cite{CBS:Sr}. In general, however, photons probe dipole
transitions with rather high
Zeeman degeneracy due to 
large total angular momentum. 
It was shown theoretically that an average over
all possible angular momentum orientations then leads to an
antisymmetric component of the atomic scattering tensor which 
reduces the interference contrast
of coherent backscattering considerably
\cite{Jonckheere,Mueller01}. 
Indeed, in the absence of an external magnetic field, photons are
scattered elastically by freely orientable atomic dipoles just
like electrons propagating in a
sample of magnetic impurities (spin-flip scattering), and in this case
interference corrections to the diffusive transport picture are cut off
very efficiently 
\cite{Mueller02}. 

Previous analytical studies of the CBS double
scattering contribution \textit{from} an infinite atomic medium
\cite{Mueller01} as well
as the case of multiple scattering \textit{inside} an infinite
medium \cite{Mueller02} 
treated an arbitrary single, isolated, degenerate atomic dipole transition. 
This description is \textit{a priori} accurate if
the hyperfine optical resonance under consideration is
sufficiently far away from neighboring transitions and 
if the laser detuning is sufficiently small. 
However, Kupriyanov and co-workers  \cite{Kupriyanov,Kupriyanov04} 
showed that the CBS enhancement factor
displays a slight asymmetry as function of the probe frequency around the
resonance that can only be accounted for by including the other
optical hyperfine transitions. In the framework of
their purely numerical calculation, they 
did not
give a qualitative explanation for the fact that the largest
asymmetry is observed in the $\hpar$ polarization channel of preserved
helicity and that the higher
enhancement is found towards the blue-detuned side of the transition. 

Extending our previous results, we develop in the present contribution an analytical theory
for multiple photon scattering 
from  
atoms with a uniform statistical distribution over the magnetic quantum numbers
inside each hyperfine level of the ground state, leaving the
possibility of an arbitrary population distribution
of the various hyperfine levels (for example a thermal one).
We rely on the decomposition of the scattered intensity
into irreducible components with respect to the rotation group. This
method has been developed some time ago in atomic physics
\cite{Omont77,Blum96}, but only today, with the advent of laser cooling
techniques, we can study subtle interference effects for the multiple
coherent scattering of quasimonochromatic radiation.  
As an application, we calculate the CBS enhancement factor
as a function of probe frequency for scattering from an infinite
half-space. This choice is not meant to describe a
realistic experiment but minimizes the purely geometrical effects of a
finite optical thickness that varies with the scattering mean-free path.  
Our approach explains sign and magnitude of the asymmetry
observed in \cite{Kupriyanov}. 
Furthermore, our theory permits to describe precisely how the effective
degeneracy can be reduced by tuning far from hyperfine-structure  or
fine-structure multiplets. This allows us to understand how 
more complicated objects such as molecules or
clusters, that involve more and more transitions, eventually scatter
light as classical objects with an optimal interference contrast.
Therefore, we are able to bridge the gap towards effective theories for mesoscopic photon 
physics in systems with classical scatterers \cite{Akkermans04a}. 

In Sec.~\ref{theory.sec}, we develop the general theory
of multiple photon scattering by hyperfine multiplets in
the independent scattering approximation. In Sec.~\ref{classical.sec}, we discuss the transition to classical scattering
properties. Sec.~\ref{elastic.sec} contains the limiting 
cases of purely elastic scattering and the calculation of CBS from
atoms with a hyperfine multiplet of overlapping
resonances. With Sec.~\ref{conclusio.sec}, we conclude the paper by
indicating possible extensions of the work. 

\section{Analytical theory}
\label{theory.sec}

\subsection{Theoretical description}
\label{model.sec}

\begin{figure}
\begin{center}
\psfrag{J}{$J$} \psfrag{J'}{$J'$}
\psfrag{F=|I-J|}{$F_1=|I-J|$}
\psfrag{F=I+J}{$F_2=I+J$}
\psfrag{F=|I-J'|}{$F_1'=|I-J'|$}
\psfrag{F=I+J'}{$F_4'=I+J'$}
\psfrag{dots}{$\vdots$}
\psfrag{omega11}{$\omega_{11}$}
\psfrag{omega0}{\large{$\omega_0$}}
\includegraphics[width=0.7\columnwidth]{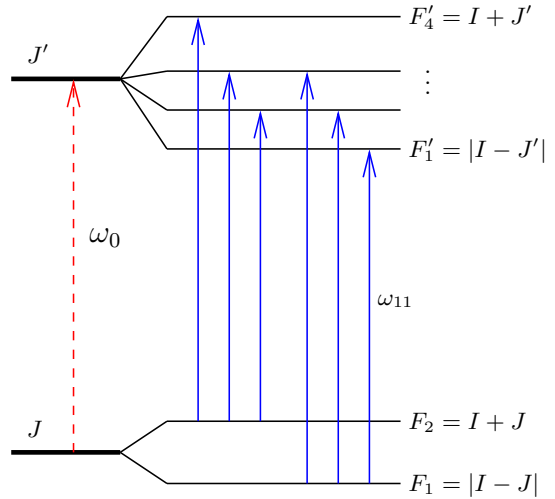}
\caption{(color online) Typical level structure of an atomic dipole transition in the
  $LS$-coupling scheme (energy splittings not drawn to scale). First the electronic spin $\bS$
  ($S=1/2$ in this example) is
  coupled to the orbital angular momentum $\bL$ (here $L=0$ in the ground
  state and $L=1$ in
the excited state) to produce the fine structure angular momentum
$\bJ$ (here $J=1/2$ and $J'=3/2$; $J''=1/2$ not shown). The
frequency of the fine structure resonance line is $\omega_0$ (dashed arrow).
$\bJ$ is then coupled to the nuclear spin
$\bI$ to give the hyperfine angular momentum
$\bF$. The allowed hyperfine dipole transitions between the levels
$\Fg$ and $\Fe'$ with frequencies $\omega_{eg}$
are marked with arrows.}
\label{multiplet.fig}
\end{center}
\end{figure}

We describe the photon field for weak laser intensity by one-photon
Fock states $\ket{\bm{k}\bm{\epsilon}}$ with wave vector $\bm{k}$ and
transverse polarization $\bm{\epsilon}$. In units where $\hbar = c =1$, the
energy or angular frequency of this state is $\omega=k$. The
scattering of these one-photon states by atomic dipole transitions
will be described following the lines of \cite{Mueller01,Mueller02}
while generalizing to multilevel transitions.  

We consider atoms at rest which are initially prepared in an electronic
ground-state with angular momentum $J$, but possibly in any
sublevel $F$ of the corresponding hyperfine multiplet. 
The probe light frequency is near-resonant with an electric dipole
transition connecting the electronic ground-state $J$ to an
electronic excited state
$J'=J-1,J,J+1$.
We assume here that the atom fine structure is well
described by an $LS$-coupling scheme (Fig.\ \ref{multiplet.fig}), where
the total electronic angular momentum
$\bJ =\bL + \bS $ couples to the nuclear spin $\bI$, producing a hyperfine multiplet
$\bF=\bJ+\bI$ with
$F=|I-J|,...,I+J$. 
The energy splitting of each hyperfine multiplet then is
much smaller than the fine structure splitting. 
 This is the
typical situation in alkali atoms like Rubidium: for the D2 line
($5\text{S}_{1/2}\to 5\text{P}_{3/2}$) of the Rb$^{85}$ isotope
with nuclear spin $I=5/2$, the ground-state $J=1/2$
is split into two levels with $F=2$ and $F=3$.
Similarly, the excited state $J'=3/2$ is split into
$F'=1,2,3,4$. 
In this scheme, the effect of different far off-resonance fine-structure
transitions $J\to J''$ with the same hyperfine quantum numbers $F \to
F'$ is neglected. 
Other coupling cases could be easily treated
along the lines of reasoning described in the present paper.
In general, several
optical transitions $F \to
F'=F-1,F,F+1$ between the ground-state
and the excited state hyperfine multiplets are allowed. We
will assume that the $J\to J'$ transition is 
\textit{closed} so that, after having scattered an incoming photon, the
atom returns to the same ground-state hyperfine multiplet but
possibly in a different hyperfine level (this is the
case for the Rb D2 line).

The Hamiltonian of this 
internal atomic structure with ground and excited levels
$g$ and $e$ is 
\eq{
H_\text{at} = \sum_g \omega_g
\Pg   +   \sum_e \omega_e' \hat P_e'
}
where $\omega_g$ is the energy of the atomic level with
angular momentum $F_g$. $\Pg$ is
the projector onto that level, 
\eq{ 
\Pg =
\sum_{\mg= - \Fg}^{\Fg}
\ket{\Fg \mg}\bra{\Fg\mg} 
} 
with a similar expression for the projector
$\Pe'$ onto excited levels with angular
momentum $\Fe'$. 
Throughout the paper, the
following convention holds: primed symbols like $\Fe'$ refer to the excited
state multiplet whereas unprimed symbols like
$\Fg$ refer to the ground-state multiplet. 

\subsection{Scattering amplitude}
\label{amplitude}

While scattering a
single photon $\ket{\bm{k} \bm{\epsilon}} \mapsto 
\ket{\tilde{\bm{k}} \tilde{\bm{\epsilon}}}$, the atom undergoes a 
transition $\Fg\to\Ff$ from an initial to a final hyperfine level. The
corresponding scattering amplitude 
$
T^{gf}(\bm{k}\bm{\epsilon} \mapsto \tilde{\bm{k}} \tilde{
\bm{\epsilon}}) = \sum_{ij} \tilde{\bm{\epsilon}}_i^* 
T_{ij}^{gf}(\omega) \bm{\epsilon}_j 
$
is a matrix element of
the scattering operator $T^{gf}(\omega)$ acting on
internal atomic states and polarization vectors. 
According to fundamental rules of quantum theory, 
the amplitudes for all indistinguishable scattering processes via 
intermediate excited states $e$ have to be added coherently:
\eqlab{ 
T^{gf}(\omega) = \sum_e T^{gef}(\omega)\ . 
} {tsum} 
The Cartesian components of the scattering operator for the partial scattering process $\Fg\to
F'_e\to \Ff$ are \cite{Mueller01}
\eqlab{
T^{gef}_{ij}(\omega) =
\frac{\sqrt{\omega\tilde{\omega}_\text{fg}}}{2\epsilon_0 V}
        \frac{\hat P_f D_i \Pe' D_j
\Pg}{\omega+\omega_g-\omega_e+\rmi\, \Gamma_{eg}(\omega)/2}
       \ . 
}{individualtante} 
Here $\bm{D}$ is the electronic dipole operator with components 
$D_i$. Because of energy
conservation the final photon frequency is 
$\tilde{\omega}_{fg}=\omega + \omega_g-\omega_f$. The
scattering is elastic when $\tilde{\omega}_{fg}=\omega$, i.e., 
when the initial and final state have the same energy $\omega_g=\omega_f$. It
is inelastic in all other instances.

The 
decay rate 
$\Gamma_{eg}(\omega)$ of each excited level is the
sum of spontaneous decay rates to all accessible final ground-states
$\Ff$: 
\eqlab{ 
\Gamma_{eg} (\omega)=
\frac{1}{3\pi\epsilon_0} \frac{1}{2F_e'+1}
\sum_{f}\tilde{\omega}_{fg}^3|\bra{\Fe'}|\bm{D}|\ket{\Ff}|^2
}{Gammaone} 
For atoms used in cold atom experiments such as alkali or
alkaline earth
atoms, the hyperfine splitting is much smaller than $\omega$ such 
that the frequency difference between the various hyperfine transitions
can be safely ignored for the evaluation of $\Gamma_{eg}(\omega)$,  which is then
a constant, the natural linewidth $\Gamma$. Note however that this approximation
can easily be relaxed: in the formulas below describing interference effects
between various hyperfine components, such as
Eq.~(\ref{sKomega_gf.eq}), one would have to use the
frequency-dependent $\Gamma_{eg}(\omega)$. 
Since the dipole operator $\bm{D}$ acts
only on the electronic angular momentum $\bm{J}$, and
not on the nuclear spin $\bm{I}$, the reduced matrix
element can then be reduced even further \cite{Edmonds60}: 
\begin{equation}
\begin{split}
\bra{(J'I)F'}|\bm{D}|\ket{(JI)F}=& (-)^{J'+I+F+1}\sqrt{(2F+1)(2F'+1)}  \\
\times &
\sixj{J'}{F'}{I}{F}{J}{1}
\bra{J'}|\bm{D}|\ket{J}\ .
\label{decomposition.eq}
\end{split}
\end{equation}
The $6j$-symbol describes the recoupling of the 4 
angular momenta
$I$, $J$, $F$, and $1$ for the vector operator $\bm{D}$. The sum
over the final hyperfine levels $\Ff$ in
\refeq{Gammaone} then can be done using a $6j$-symbol normalization
rule such that
one finds a uniform width for all excited
hyperfine levels,  
\eqlab{ 
\Gamma =
\frac{\omega_0^3 d^2 }{3\pi\epsilon_0} \ .
}{Gamma}

Here, we introduce the 
reduced  fine structure matrix element $d = \bra{J'}|
\bm{D}|\ket{J}| /\sqrt{2J'+1}$. 
For further use, we define the dimensionless
dipole operator $\bm{d}_{eg}=
\Pe' \bm{D}\Pg/d$ between the levels $\Fe$ and $\Fg$ with 
detuning
\eq{
\delta_{eg}=\omega+\omega_g-\omega_e\ .
} 
Finally, the scattering operator takes the form 
\eqlab{
T_{ij}^{gef}(\omega) =
\frac{3}{2\pi\rho(\omega)}
        \frac{\Gamma/2}{\delta_{eg}+\rmi\,\Gamma/2}
         (\bm{d}_{fe})_i (\bm{d}_{eg})_j \ .
} {individualt}
 The prefactor contains the free photon spectral density
$\rho(\omega)= V\omega^2/2\pi^2$ for a given polarization in a 
quantization volume $V$.

\subsection{Averaging over atomic degrees of freedom}
\label{average.sec}

Multiple
scattering in the dilute regime can be depicted as 
a succession of scattering events by single atoms connected by
propagation in an effective medium. 
An effective photon transport theory is obtained by a configuration average over atomic degrees of
freedom.  
The atoms are assumed to be initially prepared independently in the ground-state levels
$\Fg$ with probabilities $p_g$. These
probabilities can represent an equilibrium distribution at a certain
temperature or situations where hyperfine pumping
has been achieved. Inside each level, we assume a 
uniform statistical distribution over magnetic quantum numbers
$m_g$. Accordingly, the one-atom density matrix of internal degrees of
freedom reads 
\eqlab{ \hat \rho_\text{at} = \sum_g
p_g \; \hat\rho_g^{(0)}\ , \qquad
\hat\rho_g^{(0)} = \frac{1}{2\Fg+1} \Pg\ .
}{rhoat}
The assumption of a scalar $\hat\rho_g^{(0)}$ is reasonable for optically thick samples, where the
isotropization by multiply scattered photons dominates possible
optical pumping effects due to the incident laser. Needless to say,
this assumption greatly simplifies the calculation. In essence, an 
entirely analytical description is only manageable because the average over an isotropic
distribution like \refeq{rhoat} restores rotational invariance. 

\subsection{Scattering mean free path and total cross section}
\label{effective.sec}

The wave vector $k$ of a photon with frequency $\omega$ in the dilute
atomic medium is determined by the dispersion 
relation $k(\omega) =\omega n_\text{r}(\omega)$
where $n_\text{r}(\omega)$ is the refractive index. In scattering
media, the refractive index has an imaginary part that reflects the
fact that scattering depletes the propagating average photon field. 
This defines the scattering mean free path $\Im k(\omega) =
1/2\ell(\omega)$. 
Technically, the scattering mean free path is 
calculated via the photon self-energy, which in turn is
proportional to the 
scattering operator \refeq{tsum} averaged
over the atomic density matrix \refeq{rhoat}
\cite{Mueller01,Mueller02}.   
Carrying out the isotropic sum over $\mg$, one finds 
\eqlab{ 
\frac{1}{\ell(\omega)} = \frac{2\pi n}{k^2}\ (2J'+1)\
\sum_g{p_g\
\sum_e{\frac{(2F'_e+1)\,C_{eg}^2}{1+4\delta_{eg}^2/\Gamma^2}}} \ .
}{ell} 
Here, $n$ is the number density of atoms. The sum over the ground-state levels $g$ is due to
the simple atomic distribution \refeq{rhoat}. The sum over the
excited levels $e$ is due to the linear superposition of
scattering amplitudes, and the Lorentzian frequency
dependence comes from taking the imaginary part of the resonant denominators
 in \refeq{tsum}. 
The atomic level structure enters {via} the coefficients
$C_{eg}$, a short-hand notation for the $6j$-symbols
\eqlab{
C_{eg} =
\sixj{J'}{F_e'}{I}{\Fg}{J}{1}
}{Ceg} 
that stem from the
decomposition \refeq{decomposition}. 
Thanks to the $6j$-symbol normalization, they obey the following sum rules~\cite{Rothenberg}: 
\begin{subequations}
\label{summegsq.eq}
\begin{eqnarray}
\sum_e{(2F'_e+1)\ C_{eg}^2} &=& \frac {1}{2J+1}\ , 
\label{summegsqa.eq}\\
\sum_g{(2\Fg+1)\ C_{eg}^2} &=& \frac {1}{2J'+1} \ .
\label{summegsqb.eq}
\end{eqnarray}
\end{subequations}

Energy conservation dictates that extinction of the forward
propagating mode 
is related to the total scattering cross-section $\sigma(\omega)$ by the relation 
$\ell(\omega)^{-1} = n\sigma(\omega)$. This relation, the optical theorem
in a multiple scattering disguise, will be proved in the following
subsection.   
We can use it here to write the total scattering
cross-section as 
\begin{equation}
\sigma (\omega) =  \frac{2\pi}{k^2} (2J'+1) \sum_{g} p_g  \sum_{e}
\frac{(2F'_e+1)\, C_{eg}^2}{1+4\delta_{eg}^2/\Gamma^2} \ .
\label{sigmatot.eq}
\end{equation}
This total cross section is a weighted sum of Lorentzians of uniform 
width for all possible transitions  
$(g \to e \to f)$. 
Note that this result 
pertains to scattering from 
a single hyperfine multiplet
in both the ground and the excited states, i.e., a single
fine-structure transition $J\to J'$.  
Moreover, we consider a single dipole transition with a single
principal quantum number in the excited state such that no interference
effects between different fine-structure transitions or 
different dipole transitions with the same angular momentum (as
considered in \cite{Cardimona82})  
can appear in the total cross section.

Since the total cross section measures the total depletion
of the initial photon state, 
it includes both elastic scattering events where the
final atomic level $f$ is identical to the initial level 
$g$, but also inelastic scattering where the internal
energy of the atom is changed. A separate evaluation of elastic and
inelastic scattering is possible by considering the differential
cross section.

\subsection{Differential cross section}
\label{diffxsection.sec}

In the weak scattering approximation, the building block of
multiple scattering is the average differential
cross section for scattering polarized photons
$\ket{\bm{k} \bm{\epsilon}} \mapsto \ket{\tilde{\bm{k}}
\tilde{\bm{\epsilon}}}$ while the atom undergoes a transitions $(g \to f)$. 
The differential cross-section is obtained by averaging the square of
the scattering operator \refeq{tsum} over the atomic density matrix
\refeq{rhoat}, giving the well-known Kramers-Heisenberg formula 
\cite{Loudon}.
When all scattering processes without final frequency
analysis are considered, it can be written in the form 
\begin{equation}\label{diffsigma.eq}
\frac{\rmd\sigma}{\rmd\Omega} = 
\frac{3\sigma (\omega) }{8\pi}\
\mathcal{I}(\bm{\epsilon},\tilde{\bm{\epsilon}}^\ast,\tilde{\bm{\epsilon}},\bm{\epsilon}^\ast;\omega)
\ .
\end{equation}
Here, the total cross section $\sigma(\omega)$ has been
factorized such that the dimensionless atomic vertex
funtion $\mathcal{I}$ contains 
all angular information about light polarization and 
internal atomic structure. This vertex function has been
calculated in \cite{Mueller01} for the case of a single isolated
resonance $J\to J'$ using the techniques of irreducible tensor
operators. Since we still assume a 
\textit{scalar} density matrix \refeq{rhoat}, the vertex function 
must still be the sum of all possible scalar
products of its vector arguments, 
\eqlab{
\mathcal{I}(\bm{\epsilon},\tilde{\bm{\epsilon}}^\ast,\tilde{\bm{\epsilon}},\bm{\epsilon}^\ast ;\omega)=
        w_1 \,|\tilde{\bm{\epsilon}}^\ast\cdot\bm{\epsilon}|^2  
+ w_2 \, |\tilde{\bm{\epsilon}}\cdot\bm{\epsilon}|^2
+ w_3 \ .
}{vertexfunction} 
As in \cite{Mueller01}, the weights are given by the combinations 
\eqlab{
w_1=\frac{s_0-s_2}{3},\ 
w_2=\frac{s_2-s_1}{2},\ 
w_3=\frac{s_2+s_1}{2}}
{weights} 
of coefficients $s_K$ that describe  scalar
$(K=0)$, antisymmetric $(K=1)$, and symmetric
$(K=2)$ scattering. For the present case of multiple resonances, 
these coefficients are now
frequency dependent: 
\begin{equation}\label{sKomega.eq}
s_K(\omega)= \sum_g p_g \sum_f s_K^{gf}(\omega)
\end{equation}
where each transition $g\to f$ is described by  
\begin{equation}\label{sKomega_gf.eq}
s_K^{gf}(\omega)=  \frac{6\pi}{k^2 \sigma(\omega)} (2J'+1)^2 (2F_{f}+1)
\left| \sum_e{\frac{u_K^{egf}\, C_{eg}\,
C_{ef}}{1-2\rmi\delta_{eg}/\Gamma}}\right|^2\ .
\end{equation}
The structure of these expressions reflects well-known rules of
quantum theory: the total transition probability is the weighted
incoherent sum over initial states $g$ with occupation probability
$p_g$, and the incoherent sum over distinguishable final states $f$,
but the square
of the coherent sum of all indistinguishable amplitudes, here
related to the intermediate excited levels $e$. 
The atomic level structure enters via the coefficients 
$C_{eg}$ and 
$C_{ef}$ defined in \refeq{Ceg}. 
The essential information about the irreducible tensor modes
$K=0,1,2$ of the scattering operator is contained in the
factors 
\eqlab{u_K^{egf}=
(-1)^{\Fg-F'_e}\,
(2F'_e+1)\,
\sixj{1}{1}{K}{F_{f}}{\Fg}{F'_e}\ .
}{uKgf}
The coefficients $s_K^{gf}(\omega)$, 
together with the scattering mean
free path \refeq{ell}, are the basic ingredients for computing
multiple scattering quantities. 
The derivation of expressions \refeq{sKomega}-\refeq{uKgf} constitutes the main achievement of
the present work. From this point on, we will essentially explore its 
consequences. 

The total cross section  
\refeq{sigmatot} can be calculated from \refeq{diffsigma} 
by a sum over all final photon polarization vectors $\tilde{\bm{\epsilon}}$ and
an angular integration over the scattered photon's direction 
$\tilde{\bm{k}}/|\tilde{\bm{k}}|$. This operation on the vertex function
\refeq{vertexfunction} yields $(8\pi/3)\times (w_1+w_2+3w_3)$. 
The weights \refeq{weights} have been defined such that they obey the sum rule 
 \cite{Mueller01}
\begin{equation}\label{sumruleW.eq}
w_1(\omega)+w_2(\omega)+3w_3(\omega)=1.  
\end{equation}
which is equivalent to 
\begin{equation}\label{sumruleS.eq}
\sum_{K=0,1,2} (2K+1) s_K(\omega) = 3\ .  
\end{equation}
This last relation can be deduced from \refeq{sKomega} by virtue of
$6j$-symbol orthogonality~\cite{Rothenberg} and the sum rule
\refeq{summegsqb}. In a more general setting, this relation is shown
to be a trace-conservation property of the intensity scattering vertex 
\cite{diagvertex}. With this, we indeed recover the total cross section \refeq{sigmatot}
and thus prove the optical theorem
$\ell(\omega)=1/n\sigma(\omega)$ that was used above to derive the
total scattering cross section directly from the mean free path.     

Note that the interference between scattering amplitudes via different
excited states $e$, present in \refeq{sKomega},  disappears in the
total cross section \refeq{sigmatot}. This is a
result of the complete statistical average over the degenerate
ground-state. On the elementary level of Clebsch-Gordan coefficients,
the somewhat abstract $6j$-symbol orthogonality appears as a complete
cancellation of interference terms from equiprobable ground-states for
large detuning. 
  
\subsection{Elastic \textit{vs.} inelastic scattering} 

Expression \refeq{sKomega} of the differential cross-section
coefficients permits to distinguish between elastic scattering events
with $f=g$ and inelastic scattering events with $f\ne g$ such that we
are able to write 
\eq{
\frac{\rmd\sigma}{\rmd\Omega} = \sum_g p_g \sum_f
\left(\frac{\rmd\sigma}{\rmd\Omega} \right)
_{gf}
} 
where each transition $g\to f$ is described by its coefficients
$s_K^{gf}(\omega)$ defined in \refeq{sKomega_gf}. By
integrating over final polarization and scattering directions, we can
therefore write 
\eqlab{
\sigma(\omega) = \sigma_\text{el} (\omega)+\sigma_\text{inel}
(\omega)=  \sum_g p_g \sum_{f,e} \sigma_{gef}(\omega)
}{sigmadecomp}
with the total cross section for each elementary transition $g\to e\to
f$, 
\eqlab{
\sigma_{gef}(\omega) =  \frac{2\pi}{k^2} (2J'+1)^2 
(2\Ff+1)  \frac{(2F'_e+1)\, C_{eg}^2
C_{ef}^2}{1+4\delta_{eg}^2/\Gamma^2} . 
}{sigma_gef}
By separating elastic from inelastic contributions, we therefore find 
$\sigma(\omega) = \sigma_\text{el} (\omega) + \sigma_\text{inel}
(\omega) $ with 
\begin{eqnarray}
\sigma_\text{el} (\omega) &= & \sum_{g} p_g
\sum_{e}\sigma_{geg}(\omega), \\ 
\sigma_\text{inel} (\omega) & = &  \sum_{g} p_g \sum_{f\ne
g,e}\sigma_{gef}(\omega). 
\end{eqnarray}
In principle, a frequency analysis of the scattered photons makes it
possible to measure elastic and inelastic contributions
independently. 
Note, however, that all elastic components from the
various initial states $g$ are at the same frequency and
cannot be distinguished by a frequency analysis. 

Elastic scattering is of course desirable for coherent multiple
scattering since the scattered photons stay on resonance. Therefore, 
previous experiments of coherent backscattering from degenerate atomic
dipole transitions have been performed on various isolated hyperfine
transitions $F\to F'$ that are closed 
(see \cite{Wilkowski04} for a comprehensive list). 
Or rather, these transitions are very nearly closed: a small rate of
inelastic scattering persists because after excitation to 
an off-resonant hyperfine level,  
inelastic transitions are possible, for example the process $\Fg=3\to
\Fe'=3\to \Ff=2$ is the case of the much-studied transition $F=3\to
F'=4$ of Rb$^{85}$.  
Detuning away from the closed resonance towards the other resonances
of course increases the inelastic scattering rate. A completely
consistent theory of scattering from hyperfine multiplets therefore
needs to take into account inelastic scattering, 
without which predictions about CBS enhancement factors close to open
transitions are questionable \cite{Kupriyanov04}.   
Inelastic scattering can be incorporated in 
Monte Carlo simulations of photon trajectories but is beyond the scope
of the present article, devoted to analytical results. 
In Sec.~\ref{elastic.sec}, we will therefore specialize to the
interesting case where hyperfine multiplets
are simultaneously excited from a unique ground state which assures
purely elastic scattering.

\section{Transition to classical scattering properties}
\label{classical.sec}

The complicated radiation pattern described by
Eq.~(\ref{vertexfunction.eq}) 
is generally not the one of a classical point-dipole scatterer.
But for an isotropic atomic transition $F=0\to F'=1$, there is only
a single nondegenerate ground state and a single threefold
degenerate excited state. In such a case, the weights coefficients
in~\refeq{vertexfunction} are simply $w_1=1,w_2=w_3=0,$ and the atom scatters
light like a classical point-dipole or Rayleigh scatterer with 
\eqlab{
\frac{\rmd\sigma}{\rmd\Omega}  = \frac{3\sigma(\omega)}{8\pi}\
|\tilde{\bm{\epsilon}}^\ast\cdot\bm{\epsilon}|^2,\quad
\sigma(\omega)=\frac{6\pi}{k^2}\frac{1}{1+4\delta^2/\Gamma^2}\ .
}{classic} 
This is equivalent to 
the quasiclassical model of an elastically bound electron. 
In this case interference effects in multiple scattering are fully
preserved and the CBS enhancement factor achieves its maximal
value $2$ in the helicity-preserving channel 
\cite{CBS:Sr}. 

In optics, scattering of light---including interference effects
like CBS---by small particles (smaller than the wavelength of
light) is often successfully described by modelling the small particles
as classical point-dipole scatterers. However, such classical particles 
are very unlikely to be in a pure
$F=0$ state, and we know that atoms in  $F >  0$ states show  
very  low CBS interference in all polarization
channels \cite{CBS:At, Jonckheere}. Therefore, the validity of the classical
point-scatterer model for complex objects with possibly many internal resonances
is rather surprising. 
In this section, we therefore show how 
``classical'' coherent backscattering properties are recovered for complex
quantum objects (like atoms, molecules, or clusters) 
using a time-scale argument together with the analytical
formulation developed above.
In other
words, we close the gap between the microscopic theory for
atoms and effective models for 
classical scatterers. 

\subsection{Time scales} 

The fact that hyperfine and fine structures affect the angular distribution
of the scattered light has been recognized and understood for quite
some time~\cite{hyperfinepolar}. 
In order to grasp how coherent light scattering depends on the hyperfine
structure (or fine structure) multiplets and the detuning from resonance, it is useful
to consider the various time scales coming into play. 

Since $F$ is a good
quantum number, the semiclassical picture of hyperfine structure
is that both the nuclear spin $\bI$ and the
electronic angular momentum $\bJ$ precess around
their constant sum
$\bF=\bI+\bJ$. It
follows that the
atomic electric dipole precesses after being excited by the incoming laser field. If the laser source is very
monochromatic---linewidth smaller than the classical precession
frequency, i.e., temporal coherence longer than the precession
period---the radiated field is built by a coherent superposition
of radiating dipoles with various spatial orientations. The net
result of this coherent superposition, calculated using quantum
theory of angular momenta, is the unusual radiation pattern 
\refeq{vertexfunction}. There are, however, 
two complications: first, the atomic dipole decays because of
spontaneous emission over the time $\Gamma^{-1};$ second, if the
excitation is not exactly resonant, it also oscillates at a
frequency equal to the detuning $\delta$ from resonance.

These two effects are properly understood by considering the scattering of a
quasimonochromatic 
wave packet formed by superposition of different neighboring
frequencies around a central frequency $\omega$. Such a wave packet is not
scattered instantaneously, but after some delay $t_\text{W}(\omega)$,
known as the Wigner time delay. It is given by the derivative of
the scattering phase shift $\arctan(2\delta/\Gamma)$---see
\refeq{individualt}---with respect to frequency: 
\eqlab{
t_\text{W}(\omega) = \frac{2}{\Gamma} \; \frac{1}{1+4\delta^2/\Gamma^2}\ .
}{wigner} 
The Wigner time delay is the time scale after which the
atomic dipole induced by the incoming electromagnetic field ceases
to radiate coherently. It is maximum at resonance where it is
twice the lifetime of the atomic excited state, and decays 
towards zero away from resonance.

If the Wigner time delay $t_\text{W}(\omega)$ is longer than the period of
hyperfine precession $T_\Delta = 2\pi/\Delta$, where $\Delta$ is
the typical hyperfine splitting, then the radiating dipole
precesses during its coherence time, thus giving rise to a specific
radiation pattern. This is the case at resonance if the linewidth
$\Gamma$ is smaller than the hyperfine splitting $\Delta$, a
situation usually encountered for an isolated hyperfine component.
By contrast, if the Wigner time delay $t_\text{W}(\omega)$ is shorter than
the period of hyperfine precession $T_\Delta$, then one expects
the radiating dipole to be spatially frozen during its coherence
time. Consequently, the radiation pattern should turn into the one
associated with a classical dipole.

There are two obvious ways of reaching the classical scattering
situation 
$t_\text{W}(\omega) \ll T_\Delta$. 
On the one hand, if
the spontaneous decay rate is larger than the hyperfine splitting,
$\Gamma \gg \Delta$,  the various components
of the hyperfine multiplet are not resolved. The degree of freedom
associated with the nuclear
spin and its interaction with the electron are completely quenched on the
radiative decay time scale and can be forgotten, 
whatever the detuning from resonance. 
On the other hand, when   
the resonances are well separated, $\Gamma \ll \Delta$, one can
use large detuning to bring the Wigner time delay below the hyperfine
precession period. By detuning far away from the entire multiplet
of hyperfine resonances, the excitation then no longer
probes the nuclear spin, and therefore scattering is only sensitive to
the fine structure. 

\subsection{Analytical derivation} 

The full analytical theory developed in Sec.~\ref{theory.sec}
permits one to follow precisely how the different resonance 
contributions combine to yield classical scattering
characteristics for large
detuning.  When the
Wigner time delay is shorter than the hyperfine period, all 
resonant denominators $(\delta_{eg}+\rmi \,\Gamma/2)$ can be
taken equal to a common value $(\delta+\rmi \, \Gamma/2),$
where $ \delta=\omega-\omega_0$ is now the detuning from the fine-structure
resonance line $J\to J'$ (cf.\ Fig.~\ref{multiplet.fig}). In expression
\refeq{sigmatot}, the resonant denominator can thus be
factorized from the sum over excited states $e$. 
The sum rule \refeq{summegsqa} together with the normalization
$\sum_g p_g=1$ 
 can then be used to obtain the effective fine-structure cross section 
\begin{equation}
\sigma(\omega)= \frac{6\pi}{k^2} \;
\frac{M_{JJ'}}{1+4\delta^2/\Gamma^2} \ , 
\label{sigma_tot_unique.eq}
\end{equation}
where 
\eqlab{
M_{JJ'}= \frac{2J'+1}{3(2J+1)}
}
{degeneracy}
 is the ratio of multiplicities normalized to $M_{01}=1$. 
The cross section is now independent of the hyperfine structure, i.e.,
no longer depends on the nuclear spin $I$ and the population $p_g$ of 
the various hyperfine levels. 
Similar arguments apply to the scattering of
broadband radiation by thermal atoms 
(see especially Sec.~3.4.2.\ of \cite{Omont77}).

The radiation pattern of
the scattered photons is subject to the same transformation:  the frequency-dependent
denominator can be factorized from the sum over $e$ in
(\ref{sKomega.eq}).  
For large enough detuning, the atomic medium makes no difference
between photons scattered elastically or inelastically, such that all
contributions of final states $f$ must be added. 
The resulting
sum involves the product of three $6j$ coefficients which can be
computed using the Biedenharn-Elliott sum rule~\cite{Rothenberg}
and everything boils down to the coefficients calculated in
\cite{Mueller01}: 
\eqlab{
s_K=3(2J'+1)\sixj{1}{1}{K}{J}{J}{J'}^2 \ .}
{uniquesK}
The differential cross section then is still given by
\refeq{diffsigma}, its only frequency dependence being the single Lorentzian resonance of the total
cross section \refeq{sigma_tot_unique}.

The very same arguments apply to 
the electron spin responsible for the atomic fine structure.
Indeed, the coupling of the electronic orbital angular momentum
$\bL$ with the optically inactive electronic spin
$\bS$ produces the total electronic angular momentum
$\bJ=\bL+\bS$. Again,
when the spontaneous decay rate $\Gamma$ is larger than the
fine-structure splitting (meaning that the various components of
the fine-structure multiplet are not resolved), or when the laser
frequency is far detuned from the fine-structure multiplet such
that one has to sum over all possible excited levels $J'$,
one recovers the case of a $L\to L'$ transition,
where only the orbital properties (directly related to the charge
density response to the laser excitation) of the electrons play a role.
The formulas are simply obtained from
eqs.~(\ref{sigma_tot_unique.eq}), 
\refeq{degeneracy}, and (\ref{uniquesK.eq}) through replacement of
$(J,J')$ by $(L,L').$

If the ground state of an atom is an S state with
$L=0$, which can be optically excited only to a P
state with $L'=1$, we find $M_{LL'}=1$ and the
$s_K$ coefficients \refeq{uniquesK} simply reduce to $s_0=3,s_1=s_2=0.$ We
are back to the situation (\ref{classic.eq}) where the atom radiates
exactly like a classical point-dipole scatterer since only a
single atomic transition $0 \to 1$ is optically active.

\subsection{Molecules and complex objects} 

The previous analysis can be extended to objects
slightly more complex than atoms, but whose energy spectrum and
eigenstates can still be calculated. Let us for example consider a
diatomic molecule with rovibrational structure. If the molecular
linewidths are sufficiently small for the rotational structure to
be resolved, and if the incoming light is sufficiently
monochromatic for a single rotational line to be resonant, then the
molecule will scatter light with the specific radiation pattern of
the resonant $J\to J'$ transition. This is because
the Wigner time delay is longer than the rotational period of the
molecule. In contrast, if several $J\to J' = 
J,J\pm1$ rotational transitions have to
be taken into account coherently, a sum over excited states has to
be performed, very similar to the one in \refeq{tsum}, with the
only difference that the dipole matrix elements depend now on the
molecular quantum numbers, instead of $J$ and $F.$
These dipole elements are given in \cite{Judd} for various
possible couplings. A full discussion is beyond the scope of this
paper, but the net result is as expected: the sum simplifies
thanks to sum rules over $3j$ and $6j$ symbols and the radiation
pattern depends only on the electronic dipole transition
considered. In the most common case where it is a $\Sigma \to \Pi$
transition (the molecular equivalent of an atomic  S $\to$ P 
transition), one recovers again as a net result
the fact that the
molecule scatters light like a Rayleigh scatterer.

However, when several electronic states have to be taken into account---either in atoms or in molecules---interference between
several transitions is in principle still present.
When one considers larger objects, such as polyatomic molecules for example,
the time scales are considerably affected.
The total moment of inertia increases, making the rotation slower
and the corresponding classical rotation time longer. At the same time,
excitations of the object decay faster, making the Wigner time
delay shorter. Rotational degrees of freedom are thus quenched.
Whenever the
internal evolution time is longer than the Wigner time delay, one
expects the internal structure to be frozen and irrelevant for
light scattering properties. 
Generically, the shortest internal time scale which remains
shorter than the Wigner time delay,
will be associated with the electronic excitation, which is the
degree of freedom excited in the optical frequency range. An 
object with more than  few atoms will thus 
scatter ``classically'', i.e., following the average electronic
polarizability \cite{Berestetskii}. 
For an object much smaller than the light wavelength (such as a cluster
with, say, 1000 atoms), the multipolar expansion of the electronic
polarizability will be dominated by the dipolar term,
producing the  dipole radiation pattern 
\refeq{classic}. There may of course be cases where the dipole contribution vanishes and higher orders
dominate the radiation pattern, but such situations are certainly not generic. Finally, for objects
larger than the light wavelength, the shape of the object is resolved
and determines the light scattering properties. 
When the scatterers are anisotropic and possibly oriented, for example nematic
liquid crystals in external fields, the scattering
properties remain anisotropic and are described by a  classical anisotropic
polarizability tensor~\cite{Liquid}.

\section{Elastic scattering}
\label{elastic.sec}

The general expressions of Sec.~\ref{theory.sec} simplify
considerably in the case of a unique ground state since only
elastic scattering can occur. This is the case when there is no
hyperfine splitting (because either $I=0$ or $J=0$),
or when a closed dipole transition far from other resonances is
probed (cf.\ \refig{limits}). In such cases, the $6j$-coefficients
\refeq{Ceg} obey certain selection rules and 
are most easily calculated using the sum rule eq.~\refeq{summegsqb}
that reduces to a single term 
for the only ground level $\Fg$: 
\begin{equation}
C_{eg}^2= \frac {1}{(2J'+1)\ (2\Fg+1)}, 
\label{Ceg_unique.eq}
\end{equation}
an expression which is notably independent of the excited state(s)
$e$. In the following, we distinguish between cases where a single
excited state is relevant (and we are back to the previous 
results \cite{Mueller01}) and the new situation where several excited
states have to be considered. 

\begin{figure}
\begin{center}
\psfrag{J}{$J$} \psfrag{J'}{$J'$}
\psfrag{omega}{$\omega$} \psfrag{J0}{$J\!=\!0$}
\psfrag{Je1}{$J'\!=\!1$}
\psfrag{F=I}{$F=I$}
\psfrag{F'=I-1}{$F'\!=\!I\!-\!1$}
\psfrag{F'=I}{$F'\!=\!I$}
\psfrag{F'=I+1}{$F'\!=\!I\!+\!1$}
\psfrag{Fmin}{$F_\text{min}$}
\psfrag{Fmax}{$F_\text{max}$}
\psfrag{Femin}{$F'_\text{min}$}
\psfrag{Femax}{$F'_\text{max}$} \psfrag{(a)}{\textbf{(a)}}
\psfrag{(b)}{\textbf{(b)}} \psfrag{(c)}{\textbf{(c)}}
\psfrag{dots}{$\vdots$}
\includegraphics[width=0.90\columnwidth]{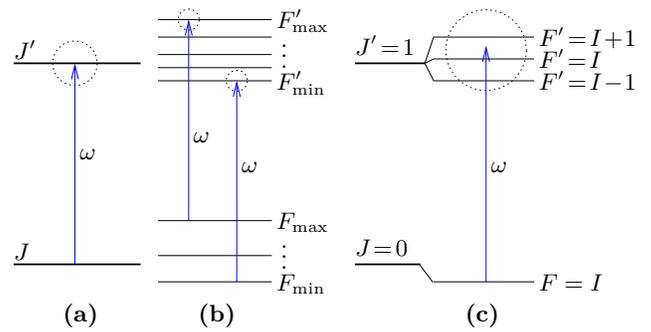}
\caption{(color online) Level scheme for limiting cases of purely elastic light
scattering, requiring a 
  unique ground-state $\Fg$. Effectively excited levels
  are enclosed by a dotted circle.  
 \textbf{(a)} Zero nuclear spin: only the fine
  structure with a unique excited state is involved. \textbf{(b)} Resonant excitation of a closed
  transition $F_{\text{max}} \to F_{\text{max}}+1$ or $F_{\text{min}}
\to F_{\text{min}}-1$, also with a unique excited level.  \textbf{(c)} $J=0$: a unique ground level with $F=I$ is coherently coupled to
several excited 
  levels. Cases (a) and (b) 
reduce to the known results of \cite{Mueller01}.  Case (c) with a
structured excited multiplet is treated analytically in section \ref{elastic.sec}.}
\label{limits.fig}
\end{center}
\end{figure}

\subsection{Unique excited state}

In the two situations shown in \refig{limits}(a) and
(b), there is only a single effective two-level system to
be considered. 
In these two cases, sums over ground and excited states reduce to a single
term and we recover the results of \cite{Mueller01}: 

$(i)$ Zero nuclear spin $I=0$. In this case, there is
no hyperfine splitting and $\Fg=J$,
$F_e'=J'$. The coupling coefficient 
\refeq{Ceg_unique} is $C_{eg}^2=1/(2J+1)(2J'+1)$.
The net effect is the same as for a large detuning from the
multiplet discussed in Sec.~\ref{classical.sec}:  we recover
the total cross section \refeq{sigma_tot_unique} and the intensity
vertex coefficients \refeq{uniquesK} for a $J\to J'$
transition.

$(ii)$ On-resonant excitation
of a \textit{closed} transition
$F_\text{max}=I+J\to F_\text{max}'=
F_\text{max} + 1 $ or
$F_\text{min}=|I-J| \to
F'_\text{min}=F_\text{min} -1 $. The latter case is
possible only if $ |I-J|\geq 1$.  Naturally, the hyperfine
structure splitting $\Delta$ should be large enough such that other resonances can
indeed be neglected.
The total
cross section  involves only a single term and is given by 
\begin{equation}
\sigma (\omega) = \frac{6\pi }{k^2} 
\frac{M_{eg}}{1+4\delta_{eg}^2/\Gamma^2}, 
\end{equation}
with the short-hand notation 
\eqlab{
M_{eg} = M_{\Fg\Fe'} = \frac{2\Fe'+1}{3(2\Fg+1)} 
}{Meg}
for the ratio of multiplicities. 
The differential cross-section coefficients \refeq{uniquesK} are 
\begin{equation}
s_K=3(2F'_e+1)\sixj{1}{1}{K}{\Fg}{\Fg}{F'_e}^2\ .
\end{equation}
These expressions are identical to eqs. (\ref{sigma_tot_unique.eq}) and
(\ref{uniquesK.eq}), with substitution of
$J$ by $\Fg$ and $J'$ by
$F'_e$ which was the situation anticipated in~\cite{Mueller01}.

\subsection{Multiple excited states}

A more interesting situation occurs when there is a unique
ground state coupled to a multiplet of excited states,
e.g., for a vanishing electronic ground-state angular
momentum $J=0$ like in \refig{limits}c. Now  the ground
level $\Fg=I$ is unique, and the excited
level is split into the hyperfine levels
$F_e'=I-1,I,I+1$ (for $I\ge 1$) or $F_e'=I,I+1$ (for $I=1/2$). Using
eq.~(\ref{Ceg_unique.eq}), the total cross section \refeq{sigmatot} can be written as
\begin{equation}
\sigma (\omega) = \frac{6\pi }{k^2}  
\sum_e \frac{ M_{eg} }{1+4\delta_{eg}^2/\Gamma^2}\ ,
\label{sigmae.eq}
\end{equation}
a sum of Lorentzians weighted by the
multiplicities \refeq{Meg} of the various excited states. More
interesting, and part of the central results of this paper, are the frequency-dependent 
intensity coefficients 
\eqlab{
s_K(\omega) = 
\frac{6\pi(2\Fg+1) }{k^2\sigma(\omega)} 
\left|\sum_e \frac{u^{egg}_K}{1 - 2\rmi\,\delta_{eg}/\Gamma}
\right|^2, 
}
{sKe}
where $u^{egg}_K$ is obtained from \refeq{uKgf} by
putting $g=f$. These coefficients permit to describe single and
multiple scattering for arbitrary detuning between and outside the
resonances. 

Expression \refeq{sKe} shows clearly that the coherent
superpositon of amplitudes \refeq{tsum} carries through to the
average intensity: the coefficients $s_K$ are squares of
interfering amplitudes and not the sum of squared amplitudes.
Therefore, the influence of other resonances may lead to subtle 
phenomena in multiple scattering, sensitive to the differential
cross section, which are not immediately visible
in the total scattering cross section \refeq{sigmae}. Indeed, by
virtue of a $6j$-symbol orthogonality, 
the total cross section is just the ``incoherent'' sum of the
individual cross sections and therefore insensitive to these
interference effects.  
In the classical picture, this can be nicely understood: the three hyperfine matrix elements actually originate from
the same optically active transition $J=0 \to J'=1$, and the precession of $\bJ$
around the nuclear spin $\bI$  modifies only its spatial repartition,
not the total scattering rate. 

In \cite{Mueller01}, elastic coherent backscattering of light by
atoms has been calculated analytically in the double scattering
approximation and for a semi-infinite scattering medium, at fixed
detuning from an isolated resonance $J\to J'$. Now,
 all results of
\cite{Mueller01} can be extended to arbitrary values of light
frequency. Similarly, using the Monte Carlo method described in
\cite{PRA2003}, the full CBS cone (with arbitrarily large
scattering orders) can be computed numerically. 

As a specific example, we will consider the case of the fermionic
 isotope $^{87}$Sr of strontium where the energy splitting between
hyperfine components is comparable to the widths of the resonances
themselves~\cite{Strontium}. \refig{sr2}(a) shows the total
scattering cross section for the $^1\text{S}_0 \to {^1\text{P}}_1$
optical transition ($\Gamma/2\pi = 32\ \text{MHz}$) with 
$\Fg=I=9/2$ and 
$F'_e= 9/2,\ 11/2,\ 7/2$.  The frequency separation between
the ($9/2 \to 9/2$) and the ($9/2 \to 11/2$) resonances is $17\
\text{MHz} =0.531\,\Gamma/2\pi$ while it is $60\ \text{MHz}
= 1.875 \, \Gamma/2\pi$ between the ($9/2 \to 9/2$) and the ($9/2
\to 7/2$) resonances. 
The vertical dotted lines
indicate the position of each hyperfine resonance which clearly
cannot be considered as isolated.

\begin{figure}
\psfrag{d/G}{\makebox[0pt][c]{detuning $\delta$ (units of $\Gamma$)}} 
\psfrag{hpar}{$\hpar$}
\psfrag{hperp}{$\hperp$} \psfrag{lpar}{$\lpar$}
\psfrag{lperp}{$\lperp$} 
\psfrag{9}{$\displaystyle\frac{9}{2}$}
\psfrag{7}{$\displaystyle\frac{7}{2}$} 
\psfrag{11}{$\displaystyle\frac{11}{2}$}
\centerline{\includegraphics[width=0.9\columnwidth]{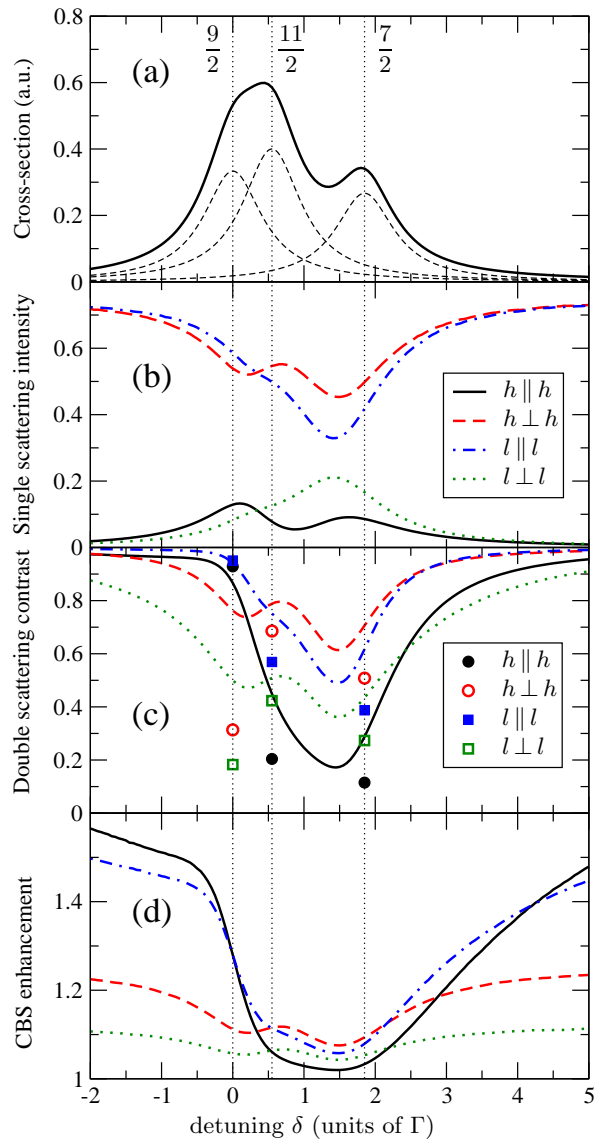}}
\caption{(Color online) \textbf{(a)} Total single scattering cross section,
  eq.~\refeq{sigmae}, for Sr$^{87}$ as a
function of detuning (in units of $\Gamma$). The ground-state
angular momentum is $\Fg=I=9/2$. There are three
accessible excited states with angular momentum
$F'_e=9/2,\, 11/2,\, 7/2$ and frequency separation
0.531$\,\Gamma/2\pi$ and 1.875$\,\Gamma/2\pi$, respectively. The
corresponding (overlapping) optical resonances are shown by dashed
lines. \textbf{(b)} Frequency dependence of the single scattering
bistatic coefficient $\gamma_1$ for a
semi-infinite medium in the four polarization
channels $\hpar$, $\hperp$, $\lpar$ and $\lperp$. 
 \textbf{(c)}
Frequency dependence of the double scattering CBS interference
contrast $\gamma_2^{(C)}/\gamma_2^{(L)}$ for the same situation. 
The predictions for
isolated resonances
\cite{Mueller01}, indicated by the symbols, are clearly off the lines,
indicating the importance of interference.
 \textbf{(d)} Frequency dependence of the total CBS enhancement factor
(\protect\ref{alpha}). Monte Carlo calculation 
including all orders of scattering for a semi-infinite medium.}
\label{sr2.fig}
\end{figure}

\subsection{Single backscattering from a semi-infinite medium}

The amount of light reflected off the sample after $N$ scattering
events can be quantified by a dimensionless parameter called the
bistatic coefficient $\gamma_N$. 
The single scattering bistatic coefficient $\gamma_1$ in the backward
direction for a semi-infinite
medium with constant spatial density exposed to an incoming plane
wave at normal incidence is essentially the atomic polarization vertex
\refeq{vertexfunction}  
(see eq.~(42) of \cite{Mueller01}, namely 
\eqlab{
\gamma_1(\omega)=  \frac{3}{4}[w_1(\omega) \,|\tilde{\bm{\epsilon}}^\ast\cdot\bm{\epsilon}|^2  
+ w_2 (\omega)\, |\tilde{\bm{\epsilon}}\cdot\bm{\epsilon}|^2
+ w_3(\omega)] \ .
}{gamma1}
In \refig{sr2}(b), $\gamma_1$ is plotted  
 as function of the detuning from the transition
$9/2 \to 9/2$ by 
using the frequency-dependent weights $w_i(\omega)$
given by \refeq{weights} in terms of (\ref{sKe.eq}). The four curves
correspond to the usual polarization channels: 
$\hpar$ (parallel
helicities, i.e., opposite circular polarizations in the backward
direction), $\hperp$ (orthogonal
helicities), $\lpar$ (parallel linear polarizations), and $\lperp$
(orthogonal linear polarizations). 
Since we take a semi-infinite medium, every photon entering the medium
must eventually exit. The atomic internal structure only redistributes
photons into different polarisation channels. The total intensity 
is independent of frequency, $\gamma_1(\lpar)+\gamma_1(\lperp) = \gamma_1(\hpar)+\gamma_1(\hperp)=3/4$.  

Far from resonance, the atom radiates like a
point-dipole scatterer, in agreement with
the general discussion of the transition to quasi-classical
scattering (section \ref{classical.sec}): since the radiated polarization is equal
to the incoming one, the bistatic coefficient vanishes in the
$\lperp$ and $\hpar$ channels and takes its maximal value 3/4 in the
$\lpar$ and $\hperp$ channels. Close to resonance, the situation is more
complex. Anisotropic scattering populates  all polarization channels and
interference effects between the various hyperfine components are
clearly made visible by the polarization analysis.

\subsection{Double scattering interference contrast} 

The multiply scattered intensity ($N>1$) contains two dominant contributions: 
first, the so-called ladder contribution
$\gamma_N^{(L)}$ without interference between multiple scattering
amplitudes. This corresponds to waves co-propagating along
scattering paths. Second, the  
so-called maximally crossed contribution $\gamma_N^{(C)}$ which
originates from counter-propagating waves and 
incorporates the interference responsible for the CBS peak. 
The CBS interference contrast at
scattering order $N$ is the ratio 
$\gamma_N^{(C)}/\gamma_N^{(L)}$. For double scattering, $N=2$, the formulas (B6) and (B24) of
\cite{Mueller01} directly give $\gamma_2^{(L)}$ and $\gamma_2^{(C)}$
at exact backscattering in terms 
of the weights $(w_1,w_2,w_3)$. 
In \refig{sr2}(c), the double scattering CBS contrast $\gamma_2^{(C)}/\gamma_2^{(L)}$ is
plotted as function of detuning.  
The on-resonant predictions for well separated resonances, shown as
symbols, neglect the interference between the different resonances
and are clearly not suited for quantitative predictions. Outside
the multiplet, the double scattering CBS interference contrast  approaches
the maximum value 1 predicted for the isotropic dipole or
non-degenerate transition $J=0\to J'=1$ in all channels 
(beyond double scattering,  the contrast is unity only in the parallel
channels $\hpar$ and $\lpar$), again in agreement with the general
discussion in Sec.~\ref{classical.sec}. 

Considering only the resonance with highest frequency, here $9/2\to
7/2$, it is evident that the interference contrast is generically larger
towards the blue side in order to reach the optimal value 1 at large
detuning. Furthermore, the largest asymmetry or slope on resonance is found for the
$\hpar$ channel since it starts from the lowest contrast on
resonance. These features, present in the figures of
\cite{Kupriyanov} but unexplained by the authors, therefore find a
natural explanation.   

\subsection{CBS enhancement} 

The enhancement factor of the CBS cone is
\begin{equation}
\label{alpha}
\alpha=1+\frac{\gamma_C}{\gamma_L+\gamma_1},
\end{equation}
where $\gamma_L=\sum_{N \ge 2} \gamma_N^{(L)}$ and $\gamma_C=\sum_{N
  \ge 2} \gamma_N^{(C)}$ account for all multiple scattering orders. 
The presently derived weights $w_i(\omega)$ can be used for evaluating
the multiply scattered intensity via the propagation eigenvalues of
the ladder and crossed series, $\lambda(\omega)$ and $\chi(\omega)$,
as derived in \cite{Mueller02}.
For third order scattering and beyond, exact analytic calculations
become very complicated, and it is more convenient to turn to a Monte Carlo
approach, as described in~\cite{PRA2003,MCRb,MonteCarlo_others}.
In \refig{sr2}(d), we plot the CBS enhancement factor for
$^{87}$Sr as a function of detuning, in the four polarization
channels and for a semi-infinite medium. The most obvious
observation is that the enhancement factors are typically small in
the region of overlapping atomic transitions, and take larger
values at large detuning, as expected from the transition to classical
scattering behavior discussed in Sec.~\ref{classical.sec}. 
High orders of scattering contribute
significantly to the ladder intensity, but only weakly to the
CBS contribution because phase coherence is rapidly lost
after the average over atomic degrees of freedom. The enhancement factor thus behaves
similarly to the double scattering contrast, only amplifying the
changes with the detuning. 

A strong reduction of the CBS enhancement factor in the vicinity of a
particular resonance may have
various origins. For example, in the $\hpar$ channel, for negative
detuning, the double scattering contrast is rather high. This is
because the dominant contribution comes from the
$9/2 \to 9/2$ transition of the $J\to J$ type with large $J$ (see~\cite{Mueller01}). The
reduction of the enhancement factor here must be attributed to single
scattering. For positive detuning, the situation is opposite:
single scattering is rather low, but also the double scattering contrast
is poor.

In the orthogonal polarization channels $\lperp$ and $\hperp$, the
multiple scattering CBS shows small interference contrast, because
ladder and CBS contributions probe different field
correlations, and thus produce a small total
enhancement factor. 
For parallel
polarization channels, $\lpar$ and $\hpar$, the enhancement factor
tends to a larger value. In the  $\hpar$ channel, the coherent and
incoherent contributions are asymptotically equal while single
scattering tends to vanish. 
For infinite detuning, we recover 
the predictions for isotropic dipole scatterers 
due to Ozrin \cite{Ozrin92}: 2.0 ($\hpar$), 1.25 ($\hperp$), 1.75
($\lpar$), 1.12 ($\lperp$). 

Note, however, that
this limit is reached very slowly: the enhancement hardly exceeds 1.5 in the
$\hpar$ channel for the largest detuning in \refig{sr2}, although it is twice as 
large as the total splitting $\Delta$ of the hyperfine multiplet. The reason for the slow recovery is simple to understand: for a semi-infinite
medium, long scattering paths contribute significantly to the CBS
cone, with all paths beyond order $N$ giving an integrated contribution
scaling like $N^{-1/2}.$ At large detuning, the $w_i$ coefficients
tend to their limiting values like  $\delta/\Delta$. This in turns
implies that a fraction $\simeq \delta/\Delta$ of the perfect
contrast is lost at each scattering event, putting an effective
cutoff $\simeq \Delta/\delta$ on the scattering orders
contributing to the CBS cone. Altogether, this implies that the
asymptotic value 2 for the enhancement factor is reached only like
$\sqrt{\delta/\Delta}$. In other words, even a
small unresolved hyperfine structure may significantly reduce the
interference contrast. This is a clear illustration that 
CBS---and any quantum interference for that matter---is very sensitive to small couplings to uncontrolled degrees of freedom.

\section{Conclusion}
\label{conclusio.sec} 

In the present paper, we have considered multiple scattering of
photons by a disordered medium of atom at rest with several internal
resonances. We have developed the analytical calculation of the total photon scattering
cross section $\sigma(\omega)$ and the differential
cross section $(\rmd\sigma/\rmd\Omega)$ as a function of the light
frequency and of the relevant angular momentum parameters defining
the internal resonances. 
We have examined under which conditions the various resonances
conspire to yield the classical 
 model of an isotropic dipole. 
As an application of the theory to multiple elastic scattering from
hyperfine multiplets, we have calculated the CBS enhancement for a
hypothetic half-space with a homogeneous average density
of atoms. 

The present paper makes two major restrictive assumptions that should be
lifted in subsequent investigations. First, 
 the results presented in Sec.~\ref{elastic.sec} for the CBS enhancement factors have
been calculated for the case of a closed transition
with a unique ground state assuring purely elastic scattering. The more general
case of inelastic scattering from open transitions can be
treated along the same lines by using the analytical theory
developed in section \ref{theory.sec}.
Second, 
real experiments with cold atoms are not performed on semi-infinite
homogeneous media, but on inhomogeneous clouds of
finite optical thickness. Their
mean free path  and thus the optical
thickness change drastically under detuning from resonance. 
This in turn modifies the relative
weight of different
scattering orders and therefore also the enhancement factor. 
Consequently, for
scattering media of finite extent, these purely geometrical effects
must be taken into account.
An efficient numerical
Monte Carlo method for this situation has been presented in
\cite{PRA2003}. The above analytical expressions for the mean free
path 
and the intensity vertex coefficients can be introduced into this
program and permit to obtain accurate CBS enhancement
factors and full peak shapes for quantitative comparison with experiments.
However, both questions are beyond the
scope of the present contribution and will be discussed elsewhere.

\begin{acknowledgments}
Ch.M. wishes to thank J. Dupont-Roc for useful
discussions. Laboratoire Kastler Brossel de l'Universit\'e Pierre et
Marie Curie et de l'\'Ecole Normale Sup\'erieure is UMR 8552 du
CNRS. CPU time on various computers has been provided by
IDRIS. The work was supported by the \textsc{Procope}
program of MAE and DAAD.   
\end{acknowledgments}

\end{document}